\documentclass[sn-nature,Numbered]{sn-jnl}

\usepackage[markup=underlined]{changes}
\usepackage{todonotes}
\usepackage{times}
\setcommentmarkup{\todo[color={authorcolor!20},size=\scriptsize]{#3: #1}}

\usepackage{graphicx}%
\usepackage{multirow}%
\usepackage{amsmath,amssymb,amsfonts}%
\usepackage{amsthm}%
\usepackage{mathrsfs}
\usepackage{calrsfs}
\usepackage{mathtools}
\usepackage[title]{appendix}%
\usepackage[x11names,dvipsnames]{xcolor}
\usepackage{textcomp}%
\usepackage{manyfoot}%
\usepackage{booktabs}%
\usepackage{algorithm}%
\usepackage{algorithmicx}%
\usepackage{algpseudocode}%
\usepackage{hyperref}%
\usepackage[utf8]{inputenc}%
\usepackage[T1]{fontenc}%
\usepackage{url}%
\usepackage{nicefrac}%
\usepackage{microtype}%
\usepackage[export]{adjustbox}
\usepackage{wrapfig,amsthm,textcomp}
\usepackage{colortbl}
\usepackage{pifont}%
\usepackage[capitalise,noabbrev,nameinlink]{cleveref}
\usepackage{adjustbox}
\usepackage{subfigure}
\usepackage{soul}
\usepackage{cancel}
\usepackage{listings}%
\usepackage{bbm}
\definecolor{good}{HTML}{74c476}
\setcitestyle{comma,open={},close={},super}
\renewcommand\refname{References and Notes}

\newcommand{\bx}{\mathbf{x}}
\newcommand{\bz}{\mathbf{z}}
\newcommand{\tbz}{\widetilde{\mathbf{z}}}
\newcommand*{\figref}[1]{\hyperref[{#1}]{Fig.~\ref*{#1}}}
\newcommand*{\tabref}[1]{\hyperref[{#1}]{Table.~\ref*{#1}}}
\renewcommand*{\eqref}[1]{\hyperref[{#1}]{equation~(\ref*{#1})}}

\begin{document}

\title[Learning multistate kinetics with a variational multistate committor network]{Learning multistate kinetics with a variational multistate committor network}

\author*[1]{\fnm{Chenyu} \sur{Tang}}\email{chenyu.tang@cnrs.fr}
\author[1]{\fnm{Cheng Giuseppe} \sur{Chen}}
\author[2,3]{\fnm{Beno\^it} \sur{Roux}}
\author*[1,4,5,6]{\fnm{Christophe} \sur{Chipot}}\email{chipot@uchicago.edu}

\affil[1]{\orgdiv{Laboratoire de Physique et Chimie Theoriques}, \orgname{Universite de Lorraine and CNRS}, \orgaddress{\city{Vandoeuvre-les-Nancy}, \country{France}}}
\affil[2]{\orgdiv{Department of Biochemistry and Molecular Biology}, \orgname{University of Chicago}, \orgaddress{\city{Chicago}, \country{USA}}}
\affil[3]{\orgdiv{Department of Chemistry}, \orgname{University of Chicago}, \orgaddress{\city{Chicago}, \country{USA}}}
\affil[4]{\orgdiv{Theoretical and Computational Biophysics Group}, \orgname{Beckman Institute for Advanced Science and Technology, University of Illinois Urbana-Champaign}, \orgaddress{\city{Urbana}, \country{USA}}}
\affil[5]{\orgdiv{Department of Physics}, \orgname{University of Illinois Urbana-Champaign}, \orgaddress{\city{Urbana}, \country{USA}}}
\affil[6]{\orgdiv{Department of Biochemistry and Molecular Biology}, \orgname{The University of Chicago}, \orgaddress{\city{Chicago}, \country{USA}}}

\abstract{
The long-time dynamics of complex molecular systems often involves rare transitions across networks of metastable states. Building on transition-path theory, which provides a rigorous framework for describing rare transitions between two metastable states, we introduce the variational multistate committor network (VMCN), a neural framework that learns the probabilities of reaching each metastable state directly from molecular simulation data. From this representation, VMCN identifies state-specific commitment, candidate transition regions and committor-consistent pathways between state pairs, and an effective kinetic network characterized by transition rates. The model is trained using finite time-lag trajectory data together with boundary conditions defined on conservative state cores. Applications to a triple-well potential, trialanine isomerization, and the $c$--ring rotation in the V$_{\rm o}$ domain of a vacuolar ATPase show that VMCN recovers metastable organization, provides committor-consistent descriptions of transition mechanisms, and estimates state-to-state kinetics. VMCN further provides diagnostics for incomplete state decompositions and enables adaptive exploration of candidate metastable states and their connecting regions. By integrating VMCN with generative committor-guided path sampling (Gen-COMPAS) for chignolin, we start from two end point structures, identify a misfolded state and a candidate folding intermediate, and we direct subsequent sampling toward the resulting multistate transition network.
}

\keywords{rare-event sampling, committor, metastability, kinetic networks, variational learning, molecular dynamics}

\maketitle

The long-time dynamics of complex molecular systems---including processes such as protein folding, ligand binding, allosteric signaling, and molecular-motor action--—involve 
rare transitions within networks of long-lived metastable states connected by multiple transition pathways.\cite{bolhuis_transition_2000,Dror2012,Peters2016,rogal_reaction_2021,Chipot2023} A deep understanding of the relationship between molecular structure and function requires more than simply identifying these metastable conformations. 
It also requires determining, for each conformation, the probability of reaching a given target state, identifying the dominant transition pathways connecting pairs of states, and quantifying the relative contributions of competing pathways to the overall dynamics.
Such a mechanistic description provides a framework for linking the microscopic conformational dynamics of a molecular system to its emergent functional behavior. 

At the simplest level, a process comprising only two long-lived metastable states can be rigorously characterized in the context of transition-path theory (TPT).\cite{Vanden-Eijnden-2010} A key ingredient of TPT is the committor probability, $q$, that a trajectory initiated from any given configuration reaches the product state before returning to the reactant state.\cite{bolhuis_reaction_2000,E2002string, Peters2006ReactionCoordinates,
berezhkovskii_committors_2019,roux_string_2021,roux2022transition} Because it encodes exactly the information needed to locate the transition state, and to construct an optimal reaction coordinate, the committor is central to rare-event theory, and a substantial body of recent work has developed string-based, variational, and neural-network methods to learn it directly from molecular-simulation data.\cite{pan_finding_2008,khoo2019solving,he_committor-consistent_2022,
chen_discovering_2023,SL23,kang2024Parrinello,MR24,MCCCRC25,
sergio_qGNN_2025,Okada2026IonDissociation, Mori2026DeepCommittor}
However, while the classic product-reactant TPT framework offers a powerful basis for computational studies, it must be expanded because most biomolecular systems intrinsically embrace multiple metastable states. For instance, the backbone of a terminally capped trialanine isomerizes among eight distinct dihedral conformations, with several conformations offering multiple accessible destinations.\cite{chen_companion_2022} It follows that the concept of committor probability must be extended to account for the kinetic competition between the different states, and resolve the sequence of transitions among all candidate destinations.

Here, we expand on these fundamental ideas to enable a characterization of multistate systems. In particular, we introduce a variational multistate committor neural network (VMCN), which extends the two-state variational committor network (VCN)\cite{he_committor-consistent_2022,chen_discovering_2023} to this multistate setting. 
The VMCN learns a single $N$-component committor\cite{Louwerse2024PRE}, ${\bf q}({\bf z})$, the entries of which approximate the probability of reaching each of the $N$ competing metastable states before any other, starting from configuration ${\bf z}$. Each configuration is, thereby, mapped to a point on the probability simplex, i.e., the natural generalization of the two-state, $[0,1]$, committor interval to $N$ competing outcomes. At the function-space level, the $N$ state-specific Dirichlet problems are separable, and their exact solutions sum to unity. Joint training with a softmax output provides a shared parameterization that enforces non-negativity and normalization throughout finite-data optimization. Boundary conditions are imposed on conservative state cores, while the transition region between them is learned from finite time-lag dynamical consistency along trajectory pairs.

The multistate committor also gives access to state-pair structure. A key ingredient in the present analysis is the pair-transition indicator, $\chi_{ij}({\bf z}) = q_i({\bf z})q_j({\bf z})$, serving as a measure of shared commitment for any pair of states, $(i,j)$. This quantity is structurally analogous to the product weighting the density of reactive trajectories in two-state TPT.\cite{bolhuis_reaction_2000,Metzner2009TPT,berezhkovskii_committors_2019,roux2022transition} Multiplication by the equilibrium density, $\pi({\bf z})\chi_{ij}({\bf z})$, yields the corresponding pair-transition density. A large $\chi_{ij}$ identifies configurations carrying appreciable commitment to both metastable states $i$ and $j$, and provides a natural starting set for constructing committor-consistent pathways. The dynamical current subsequently quantifies direct kinetic connectivity.

This multistate representation also addresses a limitation of purely discretized kinetic models. Markov state models (MSMs),\cite{Pande2010MSM, Prinz2011MSM, Schutte2011MilestoningMSM, Chodera2014MarkovStateModels} Perron-cluster cluster analysis (PCCA),\cite{Schuette1999} the variational approach for Markov processes (VAMPs),\cite{noe_variational_2013,mardt_vampnets_2018} and related transfer-operator methods are powerful tools for extracting long-time kinetics from molecular-dynamics (MD) trajectories, but their inferred macrostates can depend sensitively on the initial feature space, time lag, microstate discretization, and the clustering, or coarse-graining, procedure used to define sets of metastable states. In the present framework, state labels are treated as hypotheses about compact regions of configurational space that unambiguously belong to a given metastable state, or conservative cores. The learned multistate committor diagnoses these hypotheses, i.e., reliable cores show high commitment to their assigned state, whereas regions with persistent competition among the modeled destinations become candidates for further state analysis. Independent dynamical validation then establishes the metastability of the proposed regions.

The VMCN also interfaces naturally with the recent generative committor-guided path sampling, or Gen-COMPAS.\cite{Tang2025GenCOMPAS} Gen-COMPAS can expand the sampled configurational space from end point structures, while the VMCN provides diagnostics of the state-decomposition completeness, and proposes kinetically persistent candidate metastable states for subsequent validation. The pair indicators, $\chi_{ij}({\bf z})$, further prioritize candidate under-sampled state-pair regions for subsequent sampling. We illustrate this adaptive workflow for the mini-protein chignolin. Starting from two end points, two-state Gen-COMPAS sampling followed by high-dimensional state decomposition identifies a misfolded state, and the VMCN proposes a candidate folding intermediate. The VMCN is then trained in the full $3M-6$ internal-coordinate space, where $M$ stands for the number of atoms at hand, and configurations with large $\chi_{ij}$ are used to guide further sampling of the resulting four-state working model (i.e., unfolded, misfolded, intermediate, and folded states). The same framework can be combined with graph neural network (GNN)-style Cartesian committor learning,\cite{sergio_qGNN_2025} by replacing the molecular representation with an equivariant GNN, while retaining the multistate committor definition and training objective. More generally, the learned components, $q_i(\mathbf{z})$, describe state-specific commitments, the pair indicators, $\chi_{ij}({\bf z})$, reveal candidate pair-transition regions and initialize committor-consistent pathways, and committor-derived pairwise steady-state reactive fluxes define an effective kinetic network from which jump probabilities, direct transition rates, and mean first-passage times (MFPTs) can be computed (see Supplementary Information (SI)).

We assess the present framework examining transitions in a triple-well potential, the isomerization of a terminally capped trialanine in vacuum, the reversible folding of chignolin in water, and the $c$--ring rotation in the autoinhibited V$_{\rm o}$ domain of a vacuolar ATPase (V-ATPase).\cite{RAT05,Hovan2019,chen_companion_2022,roh2020vo,Blanc2024PNAS} Taken together, these molecular systems test multistate competition, state-decomposition refinement, adaptive discovery and sampling of previously unknown metastable states, and a predominantly sequential transition mechanism.

\section*{Results}\label{sec:results}

\subsection*{Learning the multistate committor on a triple-well potential energy landscape}

We first assess the VMCN on a two-dimensional, $(X,Y)$, triple-well potential using unbiased MD simulations. Although simple, this system involves a complication absent from a double-well potential model, namely three competing metastable
basins and three possible direct state-pair transitions. This rudimentary model system, therefore, provides a controlled test of whether a single, globally normalized multistate
committor can replace multiple pairwise committors.

\begin{figure}[t]
\centering
\includegraphics[width=\linewidth]{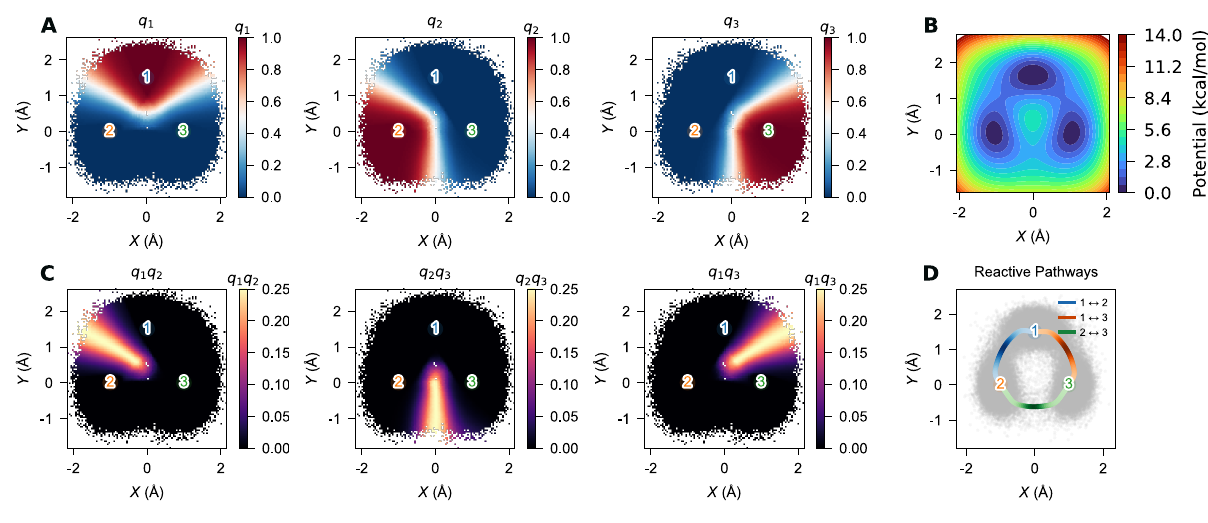}
\caption{\textbf{Triple-well potential benchmark.}  (\textbf{A}) Learned committor components, $q_i(X,Y)$, identify the commitment to each metastable basin.    
(\textbf{B}) Potential energy landscape for this triple-well potential. (\textbf{C}) Pair-transition indicators, $\chi_{ij}(X,Y)$, identifying candidate direct transition regions separating state pairs. 
(\textbf{D}) Committor-consistent pathways obtained by following the gradient of the committor, $\nabla q_i(X,Y)$.}
\label{fig:triple_well}
\end{figure}

The learned components, $q_1(X,Y)$, $q_2(X,Y)$, and $q_3(X,Y)$, correctly identify the three basins of the potential energy landscape (\figref{fig:triple_well} A,B).
Each component is close to one in its corresponding well, and close to zero in the other two wells, while the transition regions interpolate smoothly between competing destinations. The pair-transition indicators, $\chi_{ij}(X,Y)$, then localize candidate direct transition regions between each state pair (\figref{fig:triple_well} C). Because each $\chi_{ij}(X,Y)$ is large only when both commitments are appreciable, configurations already committed to either end point state are filtered out. Additional filtering by the remaining committor components can be introduced to distinguish pairwise transition tubes from multistate junctions.

Starting from high-$\chi_{ij}(X,Y)$ regions, we construct
committor-gradient-consistent pathways.\cite{chen_flow_2026} The resulting pathways connect the three state pairs, i.e., $1\leftrightarrow2$, $2\leftrightarrow3$, and $1\leftrightarrow3$ (\figref{fig:triple_well} D). This benchmark illustrates three outputs of the
method, namely (i) the components of $\mathbf{q}(X,Y)$ identify metastable commitments, (ii) the pair-transition indicators, $\chi_{ij}(X,Y)$, identify candidate pair-specific transition regions, and (iii) the committor-consistent pathways describe the corresponding reactive-network topology. These pathways provide a geometrical representation of the learned commitment field. In addition, the VMCN committors supply the direct steady-state reactive fluxes between metastable states, from which we can construct the effective kinetic generator, $\mathbf{K}$. Its off-diagonal elements are the direct transition rates, $k^{\rm direct}_{ij}$, from state
$i$ to state $j$. The VMCN-derived rates are compared with reference values obtained by direct transition counting (\tabref{tab:tw_k_direct}).

An alternate triple-well potential benchmark is reported in the SI, where the intermediate well creates an intrinsically stepwise transition mechanism. This example ascertains that the VMCN can resolve not only the end points of a reaction, but also the distinct intermediate commitment pattern associated with sequential barrier crossing.

\begin{table}[h]
\centering
\resizebox{\linewidth}{!}{%
\begin{tabular}{lllll}
\hline
Pair
& VMCN $i\rightarrow j$ (1/ps) 
& VMCN $j\rightarrow i$ (1/ps) 
& Counting ref. $i\rightarrow j$ (1/ps) 
& Counting ref. $j\rightarrow i$ (1/ps) 
\\
\hline
$1\leftrightarrow 2$ 
& $(1.25 \pm 0.11)\times 10^{-2}$ 
& $(1.74 \pm 0.12)\times 10^{-2}$ 
& $(1.39 \pm 0.04)\times 10^{-2}$ 
& $(1.94 \pm 0.05)\times 10^{-2}$ 
\\
$1\leftrightarrow 3$ 
& $(1.27 \pm 0.14)\times 10^{-2}$ 
& $(1.82 \pm 0.17)\times 10^{-2}$ 
& $(1.42 \pm 0.04)\times 10^{-2}$ 
& $(2.05 \pm 0.05)\times 10^{-2}$ 
\\
$2\leftrightarrow 3$ 
& $(3.23 \pm 0.40)\times 10^{-3}$ 
& $(3.44 \pm 0.55)\times 10^{-3}$ 
& $(3.12 \pm 0.20)\times 10^{-3}$ 
& $(3.25 \pm 0.21)\times 10^{-3}$ 
\\
\hline
\end{tabular}%
}
\caption{Rate constants, $k^\mathrm{direct}_{i\to j}$, determined from the VMCN and from counting references for the triple-well potential benchmark.} 
\label{tab:tw_k_direct}
\end{table}

\subsection*{Trialanine isomerization in vacuum }

\begin{figure}[t]
\centering
\includegraphics[width=\linewidth]{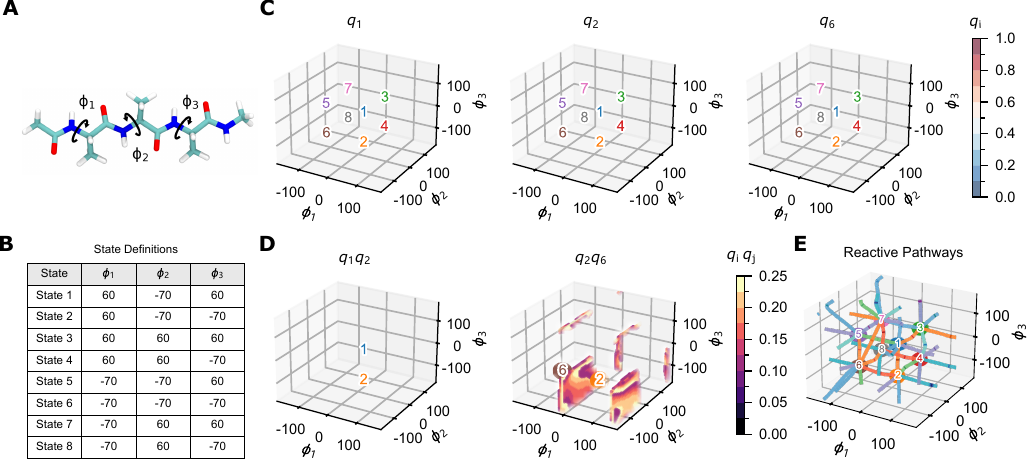}
\caption{\textbf{Multistate committor analysis of trialanine isomerization in vacuum.}
\textbf{(A)} Trialanine in vacuum and definition of the backbone dihedral angles $(\phi_1,\phi_2,\phi_3)$ used as input collective variables.
\textbf{(B)} Eight predefined metastable conformational states in the $(\phi_1,\phi_2,\phi_3)$ space.
\textbf{(C)} Representative components of the learned multistate committor, $q_1$, $q_2$, and $q_6$, showing next-hit probabilities for selected conformational states.
\textbf{(D)} Pair-reactive factors $\chi_{12}=q_1q_2$ and $\chi_{26}=q_2q_6$, which localize candidate transition regions between selected state pairs.
\textbf{(E)} Committor-consistent pathway network connecting the eight conformational states, constructed from committor-weighted dynamical exchanges inferred from the learned multistate committor $\mathbf{q}$.}
\label{fig:trialanine}
\end{figure}

We next consider trialanine isomerization in vacuum
(\figref{fig:trialanine} A).\cite{RAT05,Hovan2019} Although trialanine is small, its conformational dynamics contain several competing metastable states and numerous transition routes between them. The slow motion is commonly
described by the three backbone dihedral angles
$(\phi_1,\phi_2,\phi_3)$, the free-energy landscape of which contains eight well-characterized metastable regions.\cite{chen_companion_2022} These eight
conformational domains provide a natural test case for learning a multistate committor map and constructing a reactive kinetic network
(\figref{fig:trialanine} B).

The VMCN is trained on long unbiased trajectories using
$(\phi_1,\phi_2,\phi_3)$ as input features. The resulting commitment field provides a continuous probabilistic representation of trialanine isomerization in
dihedral space. High values of each component localize around the corresponding conformational basin, whereas transition regions display mixed commitment to competing destinations. The main outputs are summarized in \figref{fig:trialanine}, with the complete results relegated to the SI.
Representative components of the learned $\mathbf{q}$, namely $q_1$, $q_2$,
and $q_6$, show the commitment probability associated with selected conformational states (\figref{fig:trialanine} C). The corresponding pair-transition indicators, e.g., $\chi_{12}$ and $\chi_{26}$, reveal candidate direct transition regions between selected state pairs
(\figref{fig:trialanine} D). Starting from high-$\chi_{ij}$ regions, committor-gradient-consistent pathways are then constructed by following the committor gradients. Distinct pathways are subsequently compared and clustered according to their dynamical exchanges (\figref{fig:trialanine} E).\cite{chen_flow_2026}

The learned multistate committor, therefore, converts the eight-state conformational description of trialanine into a continuous transition representation and an effective kinetic model. Direct state-to-state reactive fluxes are estimated from time-lagged trajectory pairs fed to the VMCN. Together with the state populations, these steady-state reactive fluxes
define the off-diagonal generator elements, or the direct transition rates, $k^{\rm direct}_{ij}$, between metastable conformations. The resulting network identifies which state pairs are directly connected, and quantifies the relative strengths of these exchanges within a single multistate
committor-based model (see \tabref{tab:trialanine_k_direct} for representative
examples and the SI for the complete table of $k^{\rm direct}_{ij}$).

As a reference, the direct transition rates are also estimated using conventional state-to-state counting, wherein the number of observed transitions from $i$ to $j$ is normalized by the accumulated residence time at state $i$. This estimator depends on discrete-state assignments, and becomes statistically uncertain for rarely observed transitions. By contrast, the VMCN estimator uses the continuous committor field and the corresponding
reactive flux estimates. Agreement between the two computed quantities, therefore, provides a direct consistency check on the inferred kinetic network.

\begin{table}[h]
\centering
\resizebox{\linewidth}{!}{%
\begin{tabular}{lllll}
\hline
Pair
& VMCN $i\rightarrow j$ (1/ps)
& VMCN $j\rightarrow i$ (1/ps)
& Counting ref. $i\rightarrow j$ (1/ps)
& Counting ref. $j\rightarrow i$ (1/ps)
\\
\hline
$1\leftrightarrow 2$
& $(6.31 \pm 2.99)\times 10^{-6}$
& $(2.93 \pm 0.00)\times 10^{-10}$
& $(9.83 \pm 5.67)\times 10^{-6}$
& $\mathrm{N/A}$
\\
$1\leftrightarrow 6$
& $(7.79 \pm 0.00)\times 10^{-11}$
& $(1.52 \pm 0.00)\times 10^{-16}$
& $\mathrm{N/A}$
& $\mathrm{N/A}$
\\
$2\leftrightarrow 6$
& $(9.30 \pm 0.11)\times 10^{-4}$
& $(2.18 \pm 0.03)\times 10^{-4}$
& $(6.99 \pm 0.07)\times 10^{-4}$
& $(2.26 \pm 0.02)\times 10^{-4}$
\\
\hline
\end{tabular}%
}
\caption{Representative direct transition rate constants, $k^\mathrm{direct}_{i\to j}$, between states 1, 2, and 6 in \figref{fig:trialanine}, compared with direct-counting references for trialanine isomerization (N/A indicates that no corresponding direct transition events were observed in the reference trajectories).}
\label{tab:trialanine_k_direct}
\end{table}

\subsection*{Multistate committor diagnostics and adaptive Gen-COMPAS sampling}

The learned multistate committor provides a kinetic diagnostic for assessing the completeness of an initial state decomposition. State labels obtained from MSM/PCCA, free-energy calculations, or chemical intuition are treated as candidate basin cores rather than as a fixed partition. A reliable core should have high commitment to its assigned state, whereas inconsistent committor assignments indicate contaminated, or incorrectly defined cores.

Candidate missing states are identified by combining the instantaneous committor ambiguity, quantified by the normalized entropy\cite{LS22, Louwerse2024PRE} $H(\mathbf{z})$, or Gini impurity $G(\mathbf{z})$, with its finite time-lag persistence, $H_\tau(\mathbf{z})$ or $G_\tau(\mathbf{z})$. These quantities measure competition among the modeled destinations. Ordinary transition configurations can have a high instantaneous ambiguity, but rapidly commit to an existing state, whereas an omitted metastable region may retain ambiguity over the time lag, and form a coherent population with sufficient statistical support, both thermodynamically (i.e., forming a basin) and kinetically (i.e., resilient to exchanges). In the triple-well potential benchmark, this criterion distinguishes the deliberately omitted basin from the surrounding transition regions (\figref{fig:metastate}A). We explored three illustrative failure scenarios for this rudimentary system, as detailed in the SI. Long-time dynamics then validate candidate states in the molecular system.

We apply the same procedure to Gen-COMPAS trajectories of trialanine isomerization in vacuum, using the RiteWeight reweighting scheme.\cite{Tang2025GenCOMPAS,kania2026riteweight} Starting from candidate cores obtained by MSM/PCCA in the $3M-6$ heavy-atom internal-coordinate space, the VMCN removes inconsistent assignments and identifies kinetically persistent regions missing from the initial decomposition. The refined model contains eight metastable states consistent with the established trialanine conformational state decomposition (\figref{fig:metastate}B).

The VMCN framework, as well as the corresponding state-decomposition diagnostic, can be incorporated into the Gen-COMPAS workflow (\figref{fig:metastate}C). Gen-COMPAS first expands the sampled configurational space from known end points, or conservative state cores. The VMCN then refines the working state decomposition, proposes persistent candidate regions for further validation, and prioritizes under-sampled state-pair regions exhibiting a large $\chi_{ij}$. The updated state cores and selected high-$\chi_{ij}$ configurations guide the next sampling iteration, forming a closed loop between conformational sampling, candidate-state analysis, and multistate committor learning.

\subsection*{VMCN-guided Gen-COMPAS sampling of the mini-protein chignolin}

\begin{figure}[t]
\centering
\includegraphics[width=0.75\linewidth]{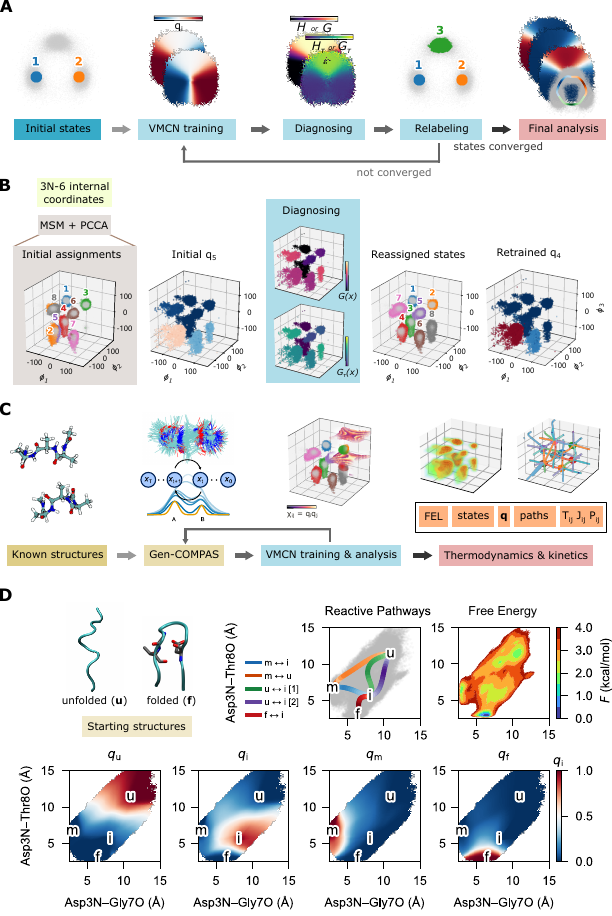}
\caption{\textbf{Multistate committor diagnostics for metastable-state refinement.}
\textbf{(A)} Schematic workflow for detecting incomplete state decompositions. A triple-well potential benchmark is initialized with one metastable state missing. Regions with high-committor uncertainty and persistent finite time-lag uncertainty are identified as candidate missing metastable states and added to the state model.
\textbf{(B)} Application to trialanine isomerization in vacuum, examined in a $3M-6$ heavy-atom internal-coordinate space. An MSM/PCCA model provides initial candidate cores, which are then screened and completed using the VMCN. The refined model recovers eight metastable states, consistent with the known trialanine conformational-state decomposition.
\textbf{(C)} A schematic of the VMCN integration with the Gen-COMPAS framework for finding metastable states and constructing a kinetic network from static end point structures. 
\textbf{(D)} VMCN-guided Gen-COMPAS candidate-state analysis and adaptive sampling for chignolin in water. The Gen-COMPAS workflow is initialized from the unfolded, $\textbf{u}$, and folded, $\textbf{f}$, states. The Asp3N--Gly7O and Asp3N--Thr8O distances are chosen for post-simulation visualization, including committor components and committor-consistent pathways connecting $\textbf{u}$, $\textbf{f}$, the misfolded state $\textbf{m}$, and the candidate intermediate, $\textbf{i}$, as well as the underlying free-energy landscape.
}
\label{fig:metastate}
\end{figure}

We illustrate the iterative workflow for chignolin folding~\cite{Honda2004Chignolin} as a common benchmark,\cite{Miao2015GaMD,Satoh2006Chignolin,Shaffer2016Chignolin,Chen2021Overcoming} using three sampling iterations (\figref{fig:metastate}D). Throughout the sampling procedure, committor modeling, state decomposition, candidate-state identification, and configuration selection are all performed in the full $3M-6$ internal-coordinate space restrained to protein carbon atoms.

The initial model contains only two end points, namely, the unfolded ($\textbf{u}$) and the folded ($\textbf{f}$) states. Iterations 0 and 1 follow the original Gen-COMPAS procedure~\cite{Tang2025GenCOMPAS}. A scalar $\textbf{u}$--$\textbf{f}$ two-state committor is trained from the accumulated trajectories and utilized to select candidate transition conformations towards further sampling. These iterations expand the explored conformational space beyond the end point basins, without assuming a single direct transition between them.

State decomposition of the accumulated conformations first resolves an additional population associated to the misfolded state, $\textbf{m}$, as documented in structural analyses of chignolin. The VMCN consistency diagnostics then identify another coherent candidate population, denoted $\textbf{i}$, and provisionally interpreted as a folding intermediate. These populations are introduced as additional state cores for the next sampling iteration, extending the original two-state description to a four-state working model containing $\textbf{u}$, $\textbf{f}$, $\textbf{m}$, and $\textbf{i}$. Additional long unbiased simulations quantify the state lifetimes and kinetic persistence.

At iteration 2, the VMCN is trained using the same high-dimensional internal-coordinate representation. Its components, $q_i(\mathbf{z})$, describe the probabilities of reaching each one of the four state cores before the others, while the pair-transition indicators, $\chi_{ij}(\mathbf{z})$, identify conformations associated with individual state-pair transitions. Conformations with large $\chi_{ij}$ are selected towards further sampling. This multistate iteration extends the end point-guided calculation to transitions involving the newly identified states.

After completing all three iterations, the combined trajectories are analyzed using the Asp3N--Gly7O and Asp3N--Thr8O distances. These simple coordinates are employed only for the final analysis, and do not enter the high-dimensional state decomposition or the sampling procedure. RiteWeight\cite{kania2026riteweight}-derived statistical weights are utilized to construct the projected free-energy landscape. A reduced VMCN is then retrained on all sampled trajectories employing these two distances as inputs and the four high-dimensional state cores as boundary labels. The resulting committor components and committor-consistent pathways provide an interpretable two-dimensional representation of the connections among the $\textbf{u}$, $\textbf{f}$, $\textbf{m}$, and $\textbf{i}$ conformational states. This example demonstrates how Gen-COMPAS and the VMCN can operate symbiotically from an end point-only description to the discovery of additional metastable states and targeted sampling of their transition network.

\clearpage

\subsection*{$c$--ring rotation in the V$_{\rm o}$ domain of a V-ATPase as a structured biomolecular transition}

We next analyze the $c$--ring rotation of the autoinhibited
V$_{\rm o}$ domain of a yeast V-ATPase. In this molecular system, rotation of the $c$--ring
relative to subunit $\rm a_{CT}$ is accompanied by a reorganization of charged and polar
contacts at the rotor--stator interface.\cite{roh2020vo,Blanc2024PNAS} Rather than seeking a
new global reaction coordinate for this process, we ask whether the multistate committor model could recover the kinetic organization of the transition from a deliberately compact structural description.

\begin{figure}[t]
\centering
\includegraphics[width=\linewidth]{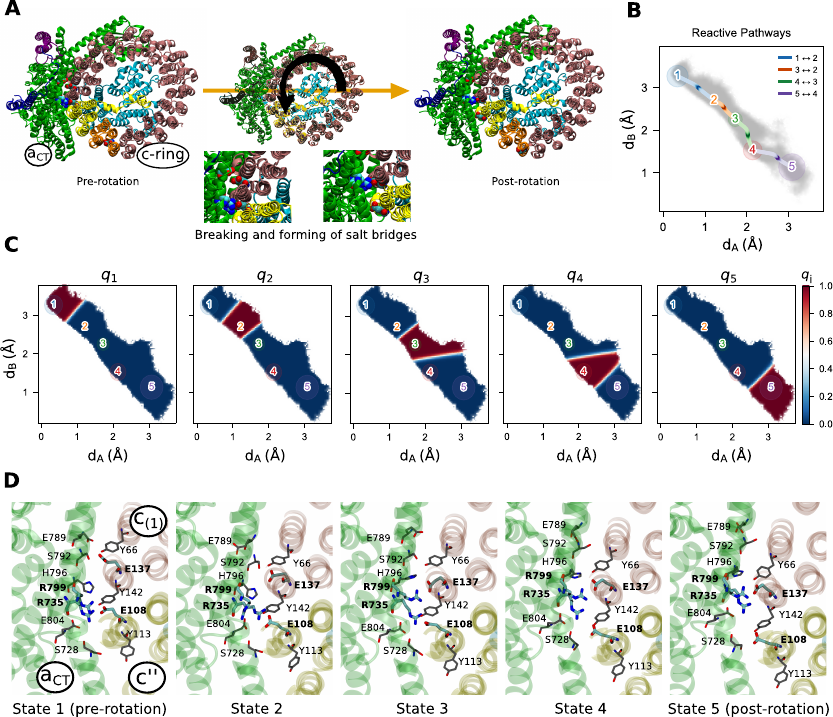}
\caption{\textbf{Multistate committor representation of the $c$--ring rotation in the autoinhibited
V$_{\mathrm{o}}$ domain of a V-ATPase.}
\textbf{(A)} V$_\mathrm{o}$ domain and its $c$--ring rotation. The CVs are defined as the distance RMSDs
with respect to the end points ($d_A$ and $d_B$).
\textbf{(B)} Committor-gradient-consistent pathways connecting the ordered states.
\textbf{(C)} Learned multistate committor components for the five states.
\textbf{(D)} Rendering of the five states and their corresponding hydrogen-bonding structures.}
\label{fig:vo_committor}
\end{figure}

We turned to the Gen-COMPAS framework\cite{Tang2025GenCOMPAS} in combination with Riteweight\cite{kania2026riteweight} to generate the trajectories and reweighting factors towards further analysis. The input features supplied to the VMCN are the distance root mean square deviations (RMSDs) with respect to the end points, $d_A$ and $d_B$, computed from the interfacial residues E108 of $\rm c''$, E137 of $\rm c(1)$, and E789, R735, and R799 of $\rm a_{\mathrm{CT}}$. These two variables monitor how the salt-bridge network changes from the initial to the final rotary configuration, without explicitly imposing a $c$--ring rotation angle. Kinetic clustering in the $(d_A,d_B)$ plane identifies five metastable states
(\figref{fig:vo_committor} A, D). Ordered from A-like to B-like, these states form a structural corridor connecting small $d_A$/large $d_B$ configurations to
large $d_A$/small $d_B$ configurations.

The learned committor components are concentrated around these five metastable regions, but vary smoothly between neighboring states (\figref{fig:vo_committor} C). Thus, the model does not just merely reproduce a geometric partition of the $(d_A,d_B)$ plane, it also assigns each configuration a set of commitment probabilities that separates the basins from the transition regions between successive rotor--stator contact networks. Based on these committor-consistent states, we further construct an effective kinetic model and estimate the transition rates, $k_{ij}^{\rm direct}$, between metastable states (\tabref{tab:vodomain_k_direct}). The resulting rate network reveals a strongly sequential mechanism for the rotary motion, with appreciable transitions occurring almost exclusively between the adjacent pairs $1\leftrightarrow2$, $2\leftrightarrow3$, $3\leftrightarrow4$, and $4\leftrightarrow5$, whereas direct transitions that bypass an intermediate state are strongly penalized. The uneven forward and backward rates between neighboring states further reveal the kinetic asymmetry of the successive steps along the rotary pathway.

The same adjacent-state topology is recovered when committor-gradient-consistent pathways are constructed from the learned multistate committor (\figref{fig:vo_committor} B). The pathways follow the curved high-density corridor in the $(d_A,d_B)$ plane, and connect the sequence $1-2-3-4-5$. This illustration is coherent with the trialanine isomerization results, brought to a larger scale for biomolecular transitions, and shows that, in the case of Gen-COMPAS sampling, two physically interpretable salt-bridge RMSDs retain enough dynamical information for the multistate committor to resolve the successive substates of the $c$--ring rotation.

\begin{table}[h]
\centering
\begin{tabular}{lll}
\hline
Pair
& VMCN $i\rightarrow j$ (1/ps) 
& VMCN $j\rightarrow i$ (1/ps) 
\\
\hline
$1\leftrightarrow 2$ 
& $(1.97 \pm 0.64)\times 10^{-4}$ 
& $(1.39 \pm 0.35)\times 10^{-4}$ 
\\
$1\leftrightarrow 3$ 
& $(1.28 \pm 1.44)\times 10^{-8}$ 
& $(7.86 \pm 3.02)\times 10^{-8}$ 
\\
$1\leftrightarrow 4$ 
& $(3.12 \pm 2.26)\times 10^{-28}$ 
& $(2.14 \pm 0.84)\times 10^{-17}$ 
\\
$1\leftrightarrow 5$ 
& $(2.12 \pm 1.52)\times 10^{-37}$ 
& $(3.27 \pm 1.52)\times 10^{-21}$ 
\\
$2\leftrightarrow 3$ 
& $(2.99 \pm 1.14)\times 10^{-5}$ 
& $(2.79 \pm 1.25)\times 10^{-5}$ 
\\
$2\leftrightarrow 4$ 
& $(4.78 \pm 2.93)\times 10^{-12}$ 
& $(1.17 \pm 0.85)\times 10^{-10}$ 
\\
$2\leftrightarrow 5$ 
& $(1.20 \pm 0.70)\times 10^{-18}$ 
& $(8.91 \pm 4.56)\times 10^{-19}$ 
\\
$3\leftrightarrow 4$ 
& $(4.88 \pm 1.96)\times 10^{-5}$ 
& $(1.95 \pm 0.55)\times 10^{-4}$ 
\\
$3\leftrightarrow 5$ 
& $(6.19 \pm 4.19)\times 10^{-8}$ 
& $(3.00 \pm 2.82)\times 10^{-6}$ 
\\
$4\leftrightarrow 5$ 
& $(4.40 \pm 2.33)\times 10^{-5}$ 
& $(7.07 \pm 3.71)\times 10^{-5}$ 
\\
\hline
\end{tabular}%
\caption{Direct transition rates, $k^\mathrm{direct}_{i\to j}$, predicted by the VMCN for the $c$--ring rotation in the V$_\mathrm{o}$ domain.}
\label{tab:vodomain_k_direct}
\end{table}

\section*{Discussion}\label{sec:discussion}

The VMCN learns a single map that assigns to each configuration a probability distribution over all labeled metastable destinations, preserving the competition among them. Its components, $q_i$, are mutually normalized first-commitment probabilities that relate metastable states through pair-transition density, $\pi q_iq_j$, and state-to-state kinetics through the direct steady-state reactive flux. The latter can be used to construct the effective generator, conditional jump probabilities, and transition rates, forming together an effective kinetic network.

The four illustrations discussed herein probe complementary aspects of the proposed framework. The triple-well potential benchmark and the isomerization of a capped trialanine in vacuum interrogate whether one normalized multistate committor can recover all competing basins and candidate pair transitions. These model systems illuminate that all predefined conformational states can be converted into a continuous commitment field, pair-specific transition regions, committor-consistent pathways, and, ultimately, an effective kinetic network. Reversible folding of chignolin illustrates the use of the VMCN diagnostics and pair-guided adaptive sampling, and the proposed multistate model is validated by extensive, unbiased sampling from the end points. For the $c$--ring rotation in the V$_{\rm o}$ domain of a yeast V-ATPase, a similar analysis reveals a sequential 
disruption and formation of salt bridges between the 
rotor and the stator, emphasizing the kinetic impossibility to bypass intermediate states.

The VMCN framework treats a state decomposition as a testable hypothesis. Own-state commitment diagnoses the purity of conservative cores, while instantaneous and time-lagged ambiguity can separate rapidly, resolving transition-state ensembles from persistent candidate metastable states. Feature and sampling controls characterize the origin of these kinetic signatures, and independent long-time simulations establish metastability. This diagnostic layer is particularly useful with end-point-driven, or adaptive sampling: Gen-COMPAS proposes intermediates and transition-state conformations, and the VMCN prioritizes their commitment, persistence, and candidate connectivity towards further simulation. Cartesian featurization through GNNs is envisioned to offer the same analytical framework directly from molecular representations.

Several assumptions, however, underlie the present interpretation. First, the coarse-graining of the conformational space, i.e., the featurizing space, must provide an adequate dynamical closure at a chosen time lag, and biased, or adaptively sampled, data require appropriate statistical weights. Second, the proposed framework rests upon the assumption of equilibrium and overdamped Langevin dynamics. Said differently, $\pi({\bf z}) q_i({\bf z}) q_j({\bf z})$ relies on equilibrium time reversibility to be interpreted as a pair-transition density. Underdamped Langevin, nonequilibrium, or irreversible dynamics, imposes a different ansatz, involving separate backward and forward multistate committors. Last, the generator inferred herein is an effective, finite-resolution kinetic model, the numerical robustness of which ought to be assessed through relevant Markovianity analyses.

\section*{Theory and Methods}\label{sec:methods}

\subsection*{From effective propagator to multistate reactive kinetics}
\label{sec:theory}

We denote by $\bx$ the molecular configuration specified by the positions of
all particles. We represent each configuration $\bx$ by a set of $d$ collective variables (CVs), 
$\bz= (z_1,\ldots,z_d)$,
via the vector-valued functions $\tilde{z}_i(\bx)$ mapping every configuration $\bx$ of the system on $\bz$. The set $\bz = \tbz(\bx)$ defines the representation
space $\Omega$, such that $\tbz(\bx)\in\Omega$. At a selected  time lag, $\tau$,
transitions from $\bz$ to $\bz'$ within $\Omega$ are described by the effective $\bz \to \bz'$ propagator $\mathcal{P}_{\tau}(\bz'|\bz)$,
\begin{equation}
\rho(\mathbf{z}';t+\tau)
=
\int_{\Omega}
\mathcal{P}_{\tau}(\mathbf{z}'  | \mathbf{z})
\rho(\mathbf{z};t)\,\mathrm d\mathbf{z}.
\label{eq:effective_propagator}
\end{equation}
We assume that the selected representation is approximately Markovian at this
time lag, that the dynamics admits a stationary density $\pi(\mathbf{z})$, and that equilibrium detailed balance
holds,\cite{roux_string_2021,roux2022transition}
\begin{equation}
\mathcal{P}_{\tau}(\mathbf{z}'|\mathbf{z})\,\pi(\mathbf{z})
=
\mathcal{P}_{\tau}(\mathbf{z}|\mathbf{z}')\,\pi(\mathbf{z}').
\label{eq:detailed_balance}
\end{equation}
Consider $N$ disjoint candidate metastable states 
$S_1,\ldots,S_N\subset\Omega$, and define the transition domain as
\begin{equation}
\Omega_0
=
\Omega\setminus\bigcup_{i=1}^{N}S_i.
\label{eq:transition_domain}
\end{equation}
For a trajectory $\mathbf{z}_t$ initiated from $\mathbf{z}_0=\mathbf{z}$,
let
\begin{equation}
T_i
=
\inf\left\{t\geq0:\mathbf{z}_t\in S_i\right\}
\label{eq:first_hitting_time}
\end{equation}
be the first hitting time of state $S_i$.
The $i$th component of the multistate committor is,
\begin{equation}
q_i(\mathbf{z})
=
\mathbb{P}_{\mathbf{z}}
\left(
T_i=\min_{1\leq k\leq N}T_k
\right),
\label{eq:multistate_committor_definition}
\end{equation}
with boundary values
$q_i(\mathbf{z})=\delta_{ij}$ for $\mathbf{z}\in S_j$. The full map is,

\begin{equation}
\left\{
\begin{array}{rll}
\mathbf{q}(\mathbf{z}) & = &
\bigl(q_1(\mathbf{z}),\ldots,q_N(\mathbf{z})\bigr)
\in\Delta^{N-1}
\\[0.4cm]
q_i(\mathbf{z}) &\geq & 0
\\[0.2cm]
\displaystyle 
\sum_{i=1}^{N}q_i(\mathbf{z}) & = & 1
\hspace*{1.0cm} 
\text{for all }\mathbf{z}\in\Omega,
\label{eq:multistate_simplex}
\end{array}
\right.
\end{equation}

\noindent
where $\Delta^{N-1}$ is referred to as the probability simplex. The normalization follows
because the next labeled core must be one of the mutually exclusive states,
$S_1,\ldots,S_N$. Consequently, $\mathbf{q}(\mathbf{z})$ defines a categorical
probability distribution over the possible next-commitment states at every
configuration. Each molecular configuration is, therefore, mapped to
barycentric coordinates over the metastable-state labels, with every
coordinate encoding a next-commitment probability.
This probabilistic normalization at every $\mathbf{z}$ also makes it possible to quantitatively express the concept of  ``local
commitment uncertainty'' using the Shannon entropy,\cite{Louwerse2024PRE}
\begin{equation}
H(\mathbf{z})
=
-\sum_{i=1}^{N}
q_i(\mathbf{z})\ln q_i(\mathbf{z}),
\qquad
0\leq H(\mathbf{z})\leq\ln N.
\label{eq:committor_entropy}
\end{equation}
The entropy vanishes when the commitment to a given metastable state is certain and reaches its maximum when
all next-commitment outcomes are equally probable.

For continuous-time reversible dynamics, each $q_i$ minimizes the generator Dirichlet energy subject to these boundary values, and satisfies the corresponding harmonic equation in the transition domain. VMCN uses the
finite time-lag approximation,
\begin{equation}
\mathcal{E}_{\tau}[\mathbf{f}]
=
\frac{1}{2\tau}
\sum_{i=1}^{N}
\left\langle
\left[
f_i(\mathbf{z}_{t+\tau})-f_i(\mathbf{z}_t)
\right]^2
\right\rangle_{\pi},
\label{eq:finite_lag_dirichlet_theory}
\end{equation}
with the simplex and core boundary
constraints.\cite{he_committor-consistent_2022,chen_discovering_2023}
Minimization of \eqref{eq:finite_lag_dirichlet_theory} gives the numerical
finite time-lag approximation used in this work and approaches the continuous-time
generator Dirichlet principle as the time lag becomes short. The committor in
\eqref{eq:multistate_committor_definition} uses continuously monitored core
hits, while the empirical loss uses time-lagged end-point pairs. Time-lag convergence
and held-out first-commitment calibration quantify this approximation.

The multistate structure gives a direct interpretation to the product
$q_i (\bz) \, q_j (\bz)$. Let $q_i^{+}(\mathbf{z})$ be the probability that the next labeled
core visited is $S_i$, and let $q_i^{-}(\mathbf{z})$ be the probability that
the most recently visited labeled core was $S_i$. Conditional on the present
configuration, the past and future are independent for a Markov process. For
equilibrium, time-reversible, overdamped dynamics represented using positions
alone,
\begin{equation}
q_i^{-}(\mathbf{z})
=
q_i^{+}(\mathbf{z})
=
q_i(\mathbf{z}).
\label{eq:forward_backward_committor}
\end{equation}
If momenta are retained in the state representation, time reversal additionally
requires momentum inversion, and the backward committor is not generally
identical to the forward committor evaluated at the same phase-space point.
Therefore, for $i\neq j$, the equilibrium density of transition-domain points
belonging to trajectories, the last visited core of which is $S_i$, and the next
visited core of which is $S_j$, can be expressed as,
\begin{equation}
\rho^{\mathrm R}_{i\to j}(\mathbf{z})
=
\pi(\mathbf{z})
q_i^{-}(\mathbf{z})
q_j^{+}(\mathbf{z})
=
\pi(\mathbf{z})
q_i(\mathbf{z})
q_j(\mathbf{z}).
\label{eq:pair_reactive_density}
\end{equation}

This multistate TPT construction motivates the introduction of the dimensionless pair-transition indicator,
\begin{equation}
\chi_{ij}(\mathbf{z})
=
q_i(\mathbf{z})q_j(\mathbf{z}).
\label{eq:chi_definition}
\end{equation}
The product reaches its maximum value $1/4$ when
$q_i=q_j=1/2$, and all other committor components vanish. Hence, high
$\chi_{ij}$ together with small
$\sum_{k\neq i,j}q_k$ identifies configurations dominated by direct
competition between states $i$ and $j$. The dimensionless factor $\chi_{ij}$
quantifies this shared commitment, and
$\pi(\mathbf{z})\chi_{ij}(\mathbf{z})$ gives the corresponding pair-transition
probability density. Dynamical transport is quantified by the  pair-reactive current, obtained by
subtracting the forward and backward contributions associated with the
observed transition $\mathbf{z}\rightarrow\mathbf{z}'$,

\begin{equation}
\begin{split}
j^{\mathrm R}_{i\to j}
(\mathbf{z},\mathbf{z}';\tau)
=
\frac{1}{\tau}
\pi(\mathbf{z})
\mathcal{P}_{\tau}(\mathbf{z}'|\mathbf{z})
\Big[
q_i(\mathbf{z})q_j(\mathbf{z}')
-
q_j(\mathbf{z})q_i(\mathbf{z}')
\Big].
\end{split}
\label{eq:pair_reactive_current}
\end{equation}

When only two metastable states are considered, $q_i=1-q_j$, and the bracket reduces to $q_j(\mathbf{z}')-q_j(\mathbf{z})$, recovering the standard two-state TPT
 edge-current factor.\cite{roux2022transition} 
For a multistate pair, it measures the change in relative
commitment to $i$ and $j$ while accounting for competition with the remaining
states. With the continuously monitored committor, this finite time-lag expression
provides a small time-lag approximation to
\eqref{eq:continuous_pair_reactive_current}. Time-lag analysis quantifies the
contribution of intervening core visits.

For a diffusion process that is Markovian in the limit of an infinitesimal time lag,\cite{roux2022transition} the continuous pair-reactive current is
\begin{equation}
 j^{\mathrm R}_{i\to j}(\mathbf z)
=
\pi(\mathbf z)D(\mathbf z)
\left[
q_i(\mathbf z)\nabla q_j(\mathbf z)
-q_j(\mathbf z)\nabla q_i(\mathbf z)
\right],
\label{eq:continuous_pair_reactive_current}
\end{equation}
where  $D(\mathbf z)$ is the diffusion tensor.
 This expression reduces to the familiar
two-state current $\pi D\nabla q_j$ when $q_i=1-q_j$.\cite{Berezhkovskii-JPC-2013}

A simple dividing surface between states $i$ and $j$ is an isocommittor surface,
\begin{equation}
\Sigma_{ij}(c)
=
\left\{
\mathbf{z}\in\Omega_0:q_j(\mathbf{z})=c
\right\},
\qquad
0<c<1.
\label{eq:isocommittor_surface}
\end{equation}
Because $q_j=0$ on $S_i$ and $q_j=1$ on $S_j$, this surface separates the two
cores. The corresponding finite time-lag edge-flux is,
\begin{equation}
J_{i\to j}(c)
=
\int_{q_j(\mathbf{z})<c}\!\mathrm d\mathbf{z}
\int_{q_j(\mathbf{z}')\geq c}\!\mathrm d\mathbf{z}'\,
j^{\mathrm R}_{i\to j}
(\mathbf{z},\mathbf{z}';\tau).
\label{eq:flux_surface}
\end{equation}
In the exact continuous-time, fully resolved theory, conservation of the reactive current makes
the flux independent of the choice of the dividing surface, provided that $S_i$ and $S_j$ lie on opposite sides, and the surface does not intersect either core. The isocommittor surface provides a convenient construct satisfying
these conditions. At finite time lag, in finite data, and in projected coordinate spaces, the estimated flux can exhibit residual surface dependence. We, therefore,
evaluate its numerical robustness as described in the SI.

The fluxes define direct kinetic observables. If $\pi_i$ is the equilibrium
population of the full metastable basin associated with state $i$ (estimated
as detailed in the SI), the direct transition rate is,
\begin{equation}
k^{\mathrm{direct}}_{i\to j}
=
\frac{J_{i\to j}}{\pi_i},
\qquad
i\neq j.
\label{eq:direct_rate}
\end{equation}
The off-diagonal elements of an effective continuous-time generator matrix, $\textbf{K}$, are,
\begin{equation}
K_{ij}
=
k^{\mathrm{direct}}_{i\to j},
\qquad
\forall i\neq j,
\end{equation}
with diagonal elements,
\begin{equation}
K_{ii}
=
-\sum_{j\neq i}K_{ij}.
\label{eq:effective_generator}
\end{equation}
Within this effective continuous-time Markov representation, the negative diagonal element $-K_{ii}$ gives the total escape rate from state $i$, and its inverse, $-1/K_{ii}$, gives the corresponding model-implied mean residence time at state $i$. Once the generator $\mathbf{K}$ has been constructed, standard kinetic observables can be obtained using established continuous-time Markov-chain analysis.\cite{Kim1958MFPT, VandenEijnden2009Milestoning, Kells2020MFPT}

\subsection*{Variational multistate committor network}

The theoretical multistate committor is defined in equations~(\ref{eq:multistate_committor_definition})--(\ref{eq:multistate_simplex}). In practice, the molecular configuration $\mathbf{x}$ is represented by a chosen feature map $\mathbf{z(x)}\in\Omega$. Depending on the application, $\mathbf{z(x)}$ may contain low-dimensional CVs, internal coordinates, or latent features produced by an invariant or equivariant GNN.

The VMCN approximates the multistate committor with a neural network, the final layer of which is a softmax,
\begin{equation}
\mathbf{q}_{\theta}(\mathbf{z})
=
\bigl(q_{\theta,1}(\mathbf{z}),\ldots,q_{\theta,N}(\mathbf{z})\bigr),
\qquad
q_{\theta,i}(\mathbf{z})\geq0,
\qquad
\sum_{i=1}^{N}q_{\theta,i}(\mathbf{z})=1.
\label{eq:nn_multistate_committor}
\end{equation}
The network is trained from time-lagged pairs $(\mathbf{z}_n,\mathbf{z}_{n+\tau})$ sampled from the simulation trajectories. With statistical weights $w_n$, the empirical finite time-lag Dirichlet loss is,
\begin{equation}
\mathcal{L}_{\mathrm{Dir}}(\theta)
=
\frac{1}{2\tau}
\frac{
\displaystyle
\sum_n w_n
\left\|
\mathbf{q}_{\theta}(\mathbf{z}_{n+\tau})
-
\mathbf{q}_{\theta}(\mathbf{z}_n)
\right\|_2^2
}{
\displaystyle
\sum_n w_n
}.
\label{eq:dirichlet_loss}
\end{equation}
For stationary reversible dynamics, this is the empirical form of \eqref{eq:finite_lag_dirichlet_theory}. In the small time-lag limit, it approaches the Dirichlet form associated with the infinitesimal generator.

The state cores enter through a boundary penalty. Let $\mathbf{e}_j$ be the one-hot vertex, i.e., a vector with a single $1$ and all other entries $0$, of the simplex associated with $S_j$. We use,
\begin{equation}
\mathcal{L}_{\mathrm{bc}}(\theta)
=
\sum_{j=1}^{N}
\mathbb{E}_{\mathbf{z}\sim\pi(\cdot| S_j)}
\left[
\left\|
\mathbf{q}_{\theta}(\mathbf{z})-\mathbf{e}_j
\right\|_2^2
\right],
\label{eq:boundary_loss}
\end{equation}
where the expectation is evaluated as a weighted average over frames labeled as members of the conservative core $S_j$. The total objective is,
\begin{equation}
\mathcal{L}(\theta)
=
\mathcal{L}_{\mathrm{Dir}}(\theta)
+
\lambda_{\mathrm{bc}}\mathcal{L}_{\mathrm{bc}}(\theta).
\label{eq:total_loss}
\end{equation}

The finite time-lag loss uses all trajectory-safe pairs from the original dynamics,
including frames in a core and pairs that enter or leave a core. Core membership
adds the one-hot boundary penalty in \eqref{eq:boundary_loss}. This construction
samples the unmodified dynamical propagator. Lag-time convergence and direct
committor validation quantify the finite time-lag approximation when a trajectory
visits a core between two recorded end points.

\subsection*{Pair-specific transition regions and committor-consistent pathways}

For each state pair $(i,j)$, the pair-transition indicator is evaluated from \eqref{eq:chi_definition}. Its maximum, $\chi_{ij}=1/4$, occurs when $q_i=q_j=1/2$ and all remaining components vanish. To suppress multistate junctions, we additionally monitor
\begin{equation}
\eta_{ij}(\mathbf{z})
=
\sum_{k\neq i,j}q_k(\mathbf{z})
=
1-q_i(\mathbf{z})-q_j(\mathbf{z}).
\label{eq:third_state_commitment}
\end{equation}
Candidate direct transition regions are selected from configurations with high $\chi_{ij}$ and small $\eta_{ij}$. The precise thresholds are chosen according to the sampling density and are tested for robustness.

Committor-consistent pathways are initialized from these pair-reactive regions. Starting from high-$\chi_{ij}$ configurations, the two branches of a pathway are propagated toward $S_i$ and $S_j$ along directions that increase the corresponding committor component. In a differentiable feature space, this is implemented using the gradients of $q_i$ and $q_j$ with respect to the input coordinates.\cite{chen_flow_2026} In an atomistic representation, the same construction can be performed through the differentiable neural featurizer. Distinct paths connecting the same state pair are compared and clustered according to their committor-weighted dynamical exchange. The resulting coordinate-dependent geometrical pathways provide a representation of increasing commitment. A diffusion tensor, or dynamical metric, extends the construction to physical current streamlines.

\subsection*{Reactive-current estimation and effective kinetic observables}

For an observed time-lagged pair $(\mathbf{z}_n,\mathbf{z}_{n+\tau})$, we evaluate the antisymmetric committor-current factor,
\begin{equation}
C_{ij}^{(n)}
=
q_i(\mathbf{z}_n)q_j(\mathbf{z}_{n+\tau})
-
q_j(\mathbf{z}_n)q_i(\mathbf{z}_{n+\tau}).
\label{eq:empirical_current_factor}
\end{equation}
To estimate the flux across the pair surface $q_{j}=c$, we define an oriented crossing indicator,
\begin{equation}
h_{ij,c}^{(n)}
=
\mathbbm{1}\!\left[q_{j}(\mathbf{z}_n)<c\right]
\mathbbm{1}\!\left[q_{j}(\mathbf{z}_{n+\tau})\geq c\right],
\label{eq:crossing_indicator}
\end{equation}
where hard indicators, $\mathbbm{1}$, may be replaced by smooth approximations. The
third-state quantity, $\eta_{ij}$, identifies surfaces crossing a pair-dominated
region, while the flux estimator integrates the complete conserved current
across the chosen surface. The empirical net forward reactive flux is,
\begin{equation}
\widehat{J}_{i\to j}(c)
=
\frac{1}{\tau}
\frac{
\displaystyle
\sum_n w_n h_{ij,c}^{(n)} C_{ij}^{(n)}
}{
\displaystyle
\sum_n w_n
}.
\label{eq:empirical_flux}
\end{equation}
In the exact theory, i.e., when the committor is ``exact'', equivalent dividing surfaces give the same steady-state reactive flux. To reduce finite-sampling and projection sensitivity, the reported direct exchange is averaged over a set $\mathcal{C}_{ij}$ of thresholds that cross the well-sampled pair-reactive tube,
\begin{equation}
\widehat{J}_{i\to j}
=
\frac{1}{|\mathcal{C}_{ij}|}
\sum_{c\in\mathcal{C}_{ij}}
\widehat{J}_{i\to j}(c).
\label{eq:averaged_flux}
\end{equation}
Statistical uncertainties in $\widehat J_{i\to j}$ are obtained from
independent trajectories, or trajectory-block resampling, with the same
statistical weights used in the flux estimator. We report this sampling
uncertainty together with the dispersion of $\widehat J_{i\to j}(c)$ across
the selected surfaces, and assess the time-lag sensitivity separately (see SI).

\subsection*{Uncertainty diagnostics and metastable state refinement}

The learned multistate committor can be used to test whether the initial state
labels are kinetically consistent. For each frame $n$ with coordinate
$\mathbf{z}_n$, we define
\begin{equation}
i^\ast(n)
=
\operatorname*{arg\,max}_{1\leq i\leq N}
q_i(\mathbf{z}_n),
\qquad
q_{\max}(n)
=
\max_{1\leq i\leq N}
q_i(\mathbf{z}_n).
\label{eq:max_committor}
\end{equation}

During state-diagnostic and relabeling runs, the boundary-loss weight, $\lambda_{\rm bc}$, is annealed, so that the initial core labels act as soft anchors rather than rigid constraints. At later stages of the diagnostic optimization, dynamical consistency dominates the objective, allowing kinetically inconsistent candidate cores to develop reduced own-label commitment despite a finite residual boundary penalty. Therefore, for a labeled frame with assigned label $y(n)$, we can define the own-label consistency as,
\begin{equation}
\ell(n)
=
q_{y(n)}(\mathbf{z}_n).
\label{eq:own_label_consistency}
\end{equation}
A reliable conservative core should have high own-label consistency for most
of its frames. A core is flagged for further inspection when a substantial
fraction of its frames satisfies,
\begin{equation}
\ell(n)<\epsilon_\ell
\qquad\text{or}\qquad
i^\ast(n)\neq y(n),
\label{eq:core_inconsistency_criterion}
\end{equation}
where $\epsilon_\ell$ is an empirically prescribed consistency threshold. Such behavior indicates that the labeled core may be too broad, internally heterogeneous, or
inconsistent with the learned kinetic partition.

We quantify uncertainty in the multistate committor using normalized entropy\cite{Louwerse2024PRE,LS22} and
normalized Gini impurity. The normalized entropy is,
\begin{equation}
H(\mathbf{z})
=
-\frac{1}{\log N}
\sum_{i=1}^{N}
q_i(\mathbf{z})
\log q_i(\mathbf{z}),
\label{eq:entropy}
\end{equation}
and the normalized Gini impurity is
\begin{equation}
G(\mathbf{z})
=
\frac{
\displaystyle
1-\sum_{i=1}^{N}q_i(\mathbf{z})^2
}{
\displaystyle
1-\frac{1}{N}
}.
\label{eq:gini}
\end{equation}
Both quantities are close to zero when one committor component dominates, and
close to one when the commitment probabilities are broadly distributed.

Candidate metastable regions combine high instantaneous uncertainty with
finite time-lag persistence, whereas ordinary transition shells relax toward a
modeled destination. We, therefore, evaluate the time-lagged entropy,
\begin{equation}
H_\tau(\mathbf{z})
=
\mathbb{E}
\left[
H(X_{t+\tau})
|
X_t=\mathbf{z}
\right]
=
\int_{\Omega}
\mathrm d\mathbf{y} \,
\mathcal{P}_\tau(\mathbf{y}|\mathbf{z})
H(\mathbf{y}),
\label{eq:lagged_entropy}
\end{equation}
and the time-lagged Gini impurity is defined analogously as
\begin{equation}
G_\tau(\mathbf{z})
=
\mathbb{E}
\left[
G(X_{t+\tau})
|
X_t=\mathbf{z}
\right]
=
\int_{\Omega}
\mathrm d\mathbf{y} \,
\mathcal{P}_\tau(\mathbf{y}|\mathbf{z})
G(\mathbf{y}).
\label{eq:lagged_gini}
\end{equation}

In practice, $H_\tau$ and $G_\tau$ are estimated directly from empirical
time-lagged transition statistics. The uncertainty of the observed successor configurations,
$\mathbf{z}_{t+\tau}$, is averaged over trajectory frames, the initial
configurations of which, $\mathbf{z}_t$, belong to the same local region, cluster, or
discretized neighborhood. The continuous integral expressions above, therefore,
represent the idealized population quantities, whereas the implementation
uses weighted discrete averages over the available time-lagged pairs.

A subset $\mathcal{R}\subset\Omega_0$ is identified as a candidate metastable region missing only when it simultaneously satisfies the following
conditions:

\begin{equation}
\left\{
\begin{array}{rllll}
H(\mathbf{z}) & > & \epsilon_H
& \text{or}\quad
G(\mathbf{z})>\epsilon_G,
&
\forall \mathbf{z}\in\mathcal{R},
\\[0.4cm]
H_\tau(\mathbf{z}) & > & \epsilon_{H,\tau}
& \text{or}\quad
G_\tau(\mathbf{z})>\epsilon_{G,\tau},
&
\forall \mathbf{z}\in\mathcal{R},
\\[0.4cm]
\displaystyle 
\int_{\mathcal{R}}
\pi(\mathbf{z})\,\mathrm d\mathbf{z}
& > &
\epsilon_\pi,
&
\end{array}
\right.
\end{equation}

where the $\epsilon$ values are empirically prescribed thresholds. The first condition
requires high instantaneous committor uncertainty. The second requires this
uncertainty to persist over the time lag. The third requires sufficient
equilibrium probability or weighted statistical support.

A conventional transition-shell configuration may satisfy the first
condition, but it should typically evolve toward one of the known basins, and,
therefore, exhibit lower time-lagged uncertainty. A missing metastable region is instead expected to retain high uncertainty over the time lag, and to form a
kinetically persistent, statistically supported component.

These diagnostics are heuristic in nature, because the multistate committor, stationary density, and time-lag transition statistics are all estimated from finite-length
simulation data. They are, therefore, used as guides for metastable state refinement
rather than as rigorous state definitions.

State refinement is applied conservatively. A labeled core may be trimmed when
a substantial fraction of its frames has low own-label consistency or high
uncertainty. A frame may be reassigned to another existing state only when (i) 
$i^\ast(n)$ differs from its assigned label, (ii) $q_{\max}(n)$ is high, and (iii) its
committor uncertainty is low. Missing candidate  states are selected from
regions with high instantaneous and time-lagged uncertainty, sufficient equilibrium statistical support, and coherent spatial or kinetic connectivity. These
regions are subsequently clustered and filtered according to their weighted density and kinetic persistence.

Two candidate regions may be merged when time-lagged transitions indicate rapid exchange between them, whereas an existing state may be split when distinct high-confidence sub-regions are separated by a kinetically persistent
low-confidence barrier. The multistate committor diagnostics provide quantitative guidance for these operations, but the final decisions to prune, reassign,
merge, or split metastable states are made conservatively and may involve human inspection of representative structures, trajectory connectivity, and the sensitivity of the results to the selected thresholds and time lag.

\section*{Code availability}

VMCN framework will be available as an installable open-source package when published. 

\section*{Acknowledgments}

C.C. is supported by the European Research Council (project 101097272 ``MilliInMicro''). 
B.R. is supported by the National Institutes of Health via grant R35-GM152124 and the National Science Foundation via grand MCB-2309048. The authors are indebted to Mayank Prakash Pandey, Radu Alexander Talmazan, and Alberto Meg{\'\i}as for fruitful discussions.

\section*{Author contributions}
Data curation: CT. Methodology and conceptualization: CT, BR, CC. Software: CT. Writing---original draft: CT.
Writing---review \& editing: CT, CGC, BR, CC. Supervision: CC.

\bibliography{sn-bibliography}

\end{document}